# Spin-orbit quantum impurity in a topological kagome magnet

Jia-Xin Yin[1*†], Nana Shumiya[1*], Yuxiao Jiang[1*], Huibin Zhou[2*], Gennevieve Macam[3*], Songtian S. Zhang[1], Hano Omar Mohammad Sura[4], Zijia Cheng[1], Zurab Guguchia[1,5], Yangmu Li[6], Qi Wang[7], Maksim Litskevich[1], Ilya Belopolski[1], Xian Yang[1], Tyler A. Cochran[1], Guoqing Chang[1], Qi Zhang[1], Brian M. Andersen[4], Zhi-Quan Huang[3], Feng-Chuan Chuang[3], Hsin Lin[8], Hechang Lei[7], Ziqiang Wang[9], Shuang Jia[2], and M. Zahid Hasan[†1,10]

[1]Laboratory for Topological Quantum Matter and Advanced Spectroscopy (B7), Department of Physics, Princeton University, Princeton, New Jersey 08544, USA.

[2]International Center for Quantum Materials and School of Physics, Peking University 100193, Beijing, China.

[3]Department of Physics, National Sun Yat-Sen University, Kaohsiung 80424, Taiwan.

[4]Niels Bohr Institute, University of Copenhagen, Universitetsparken 5, DK-2100 Copenhagen, Denmark.
[5]Laboratory for Muon Spin Spectroscopy, Paul Scherrer Institute, CH-5232 Villigen PSI, Switzerland.

[6]Condensed Matter Physics and Materials Science Division, Brookhaven National Laboratory, Upton, New York 11973, USA.

[7]Department of Physics and Beijing Key Laboratory of Opto-electronic Functional Materials & Micro-nano Devices, Renmin University of China, Beijing 100872, China.

[8]Institute of Physics, Academia Sinica, Taipei 11529, Taiwan.
[9]Department of Physics, Boston College, Chestnut Hill 02467, MA, USA.

[10]Materials Sciences Division, Lawrence Berkeley National Laboratory, Berkeley, CA 94720, USA.

†Corresponding authors, E-mail: mzhasan@princeton.edu; jiaxiny@princeton.edu

*These authors contributed equally to this work.

**Quantum states induced by single-atomic-impurities are the current frontier of material and information science[1-8]. Recently the spin-orbit coupled correlated kagome magnets are emerging as a new class of topological quantum materials[9-18], although the effect of single-atomic impurities remains unexplored. Here we use state-of-the-art scanning tunneling microscopy/spectroscopy (STM/S) to study the atomic indium impurity in a topological kagome magnet $Co_3Sn_2S_2$, which is designed to support the spin-orbit quantum state. We find each impurity features a strongly localized bound state. Our systematic magnetization-polarized tunneling probe reveals its spin-down polarized nature with an unusual moment of -5$\mu_B$, indicative of additional orbital magnetization. As the separation between two impurities progressively shrinks, their respective bound states interact and form quantized molecular orbital states. The molecular orbital of three neighboring impurities further exhibits an intriguing splitting owing to the combination of geometry, magnetism, and spin-orbit coupling, analogous to the splitting of the topological Weyl fermion line[12,19]. Our work demonstrates the quantum-level interplay between magnetism and spin-orbit coupling at an individual atomic impurity, which provides insights into the emergent impurity behavior in a topological kagome magnet and the potential of spin-orbit quantum impurities for information science.**

Understanding how impurities interact with a quantum environment is an important problem with widespread implications in physics and technology[1-8]. The atomic impurity induced state in a quantum material can uncover the nature of exotic ground states, elucidate details of electronic correlations, and produce topological excitations that cannot be natively found in systems without atomic impurities[1-8]. Being a local density of states probe, STM/S has played a key role in this research area. For instance, the Zn impurity state in a high-temperature superconductor uncovers the *d*-wave symmetry[3], the Mn impurity state in a semiconductor elucidates their ferromagnetic coupling[4], and the Fe impurity in a superconductor with topological surface states creates Majorana-like state[5]. Moreover, the single-atom induced quantum states are of tremendous value for the quantum information science, with examples including the single-atom transistor[6], atomic quantum dots[7] and the single-atom memory[8]. Recently, the spin-orbit coupled kagome magnets have emerged as a new class of quantum materials[9-16]. These materials often exhibit anomalous transport behaviors, correlated topological electronic structure, and giant spin-orbit tunability, which are often linked to the fundamental kagome band structure with magnetic topological phases. Accordingly, this family serves as a fertile platform for exploring the interplay between single atomic impurity and exotic quantum electronic structure. In our systematic exploration in this family, we find that the In impurities in $Co_3Sn_2S_2$ are unique nonmagnetic impurities that introduce a highly unusual magnetic quantum state, which not only serves as a local probe of the correlated spin-orbit coupled kagome electronic structure but also demonstrates a new type of single-atom impurity which may be useful in the development of quantum information science.

$Co_3Sn_2S_2$ has a layered crystal structure with space group $R\bar{3}m$ and hexagonal lattice constants[20] $a$ = 5.3 Å and $c$ = 13.2 Å (Fig. 1**a** and **b**). The material has a ferromagnetic ground state (Curie temperature, $T_C$ = 170K) with the magnetization arising mainly from the Co kagome lattice and is aligned along the *c* axis. The large electron negativity difference leads to the strongest bonds forming between the Co and S atoms[20]. Cleaving preferentially breaks the S-Sn bond, which leads to terminations of the S surface and the Sn surface (Fig. 1**c**). Previous STM studies of $Co_3Sn_2S_2$ have dominantly observed two surfaces, one with largely vacancy defects and the other with adatom defects[13,15,17], although definitive atomic assignment of the termination surface has been challenging. Several factors complicate this assignment, as the two surfaces have identical hexagonal lattice symmetries, the interlayer distance between the Sn and S layers are sub-Å in scale, and the constant current topographic image convolutes the spatial variation of the integrated local density of states and the geometrical corrugations[21]. Decisive experimental evidence for surface identification can be found by imaging the monolayer surface boundary and the layer-selective chemical dopants[21-23]. In this material, as shown in Fig. 1**c**, the Sn surface is above the S surface when they meet at a step edge. This evidence is directly provided in Fig. 1**d**, where we can observe the vacancy surface, lies directly above the adatom surface, allowing us to assign them as the Sn and S surfaces respectively. This is therefore consistent with the native defects on the two surfaces being Sn vacancies and Sn adatoms which were created upon cleaving. We further conclude this assignment by doping the bulk $Co_3Sn_2S_2$ single crystals with 1% In impurities which is known to substitute Sn atoms[20]. Indeed, on the surface previously identified to be Sn, we find dilute nonnative substitutional atoms with concentration consistent with the nominal In doping, suggesting these impurities to be the substituted In atoms (Fig. 1**e**).

Having determined the chemical nature of the surfaces and impurities, we now perform an extensive study on its electronic properties. We find that each impurity repeatedly features a sharp state at the energy of -267meV as shown in Fig. 2**a**. First-principles calculations show each In impurity to introduce a strong resonance, very similar to the experimental data (Fig. 2**b**). Furthermore, this impurity resonance arises from a spin-down state, opposite to the bulk magnetization which is the spin majority in this energy range. The resonance resides within the bandgap of the spin-down states, analogous to previously observed impurity resonances inside the semiconducting gap[4] and the superconducting gap[3,5]. To explore its detailed real-space feature, we systematically probe the local electronic structure for an isolated In impurity, as shown in Fig. 2**c**. The corresponding dI/dV map at the impurity resonance energy in Fig. 2**d** shows a localized pattern bound to the In impurity site. This impurity bound state couples with three neighboring Co atoms in the underlying kagome lattice, as illustrated in Fig. 2**e**. Figure 2**f** shows the representative dI/dV curves measured with increasing distance from the impurity, demonstrating the bound state decaying in intensity without detectable energy dispersion or splitting. An exponential fit to the decay yields a characteristic length scale of 2.8Å (inset of Fig. 2**f**).

To probe the magnetic nature of the impurity state, we perform tunneling experiments with a spin-polarized Ni tip under weak magnetic fields[24-27]. The bulk crystal has a coercive field[12] $B_C$~0.3T, and bulk Ni tip is a soft magnet with a $B_C \ll 0.1$T that can be easily flipped by reversing the magnetic field[27]. We measure the tunneling signal of the impurity state while sequentially applying fields along the c-axis of +0.5T, +0.1T, -0.1T, -0.5T, -0.1T, and +0.1T to systematically flip the magnetization of the tip and sample (Fig. 3**a**). This sequence allows us to perform spin-polarized measurements of the impurity, as we now explain. The +0.5T field polarizes both the sample and tip, aligning the spin of the tip and anti-aligning the spin of the impurity state, due to the spin-down nature of the impurity state. A +0.1T field does not change the polarization of either the tip or impurity. Flipping the field to -0.1T also flips the spin of the tip, leaving the spin of the impurity state unchanged (down). Here, with both tip and impurity state spins aligned down, we observe an intensity increment of the tunneling signal. The dI/dV map at the impurity energy in the inset of Fig. 3**a** also confirms the enhancement of the tunneling signal at the impurity site, which indicates the moments of the tip and the impurity are aligned with each other at -0.1T. Next, we further decrease the field to -0.5T which flips up the spin impurity state, with a corresponding reduction of the tunneling signal. Lastly, by sequentially changing the field to -0.1T and +0.1T, we flip the spin of the tip (down) and again observe an increase in the intensity. Our systematic field manipulation strongly supports that impurity state features spin-down polarization, consistent with the first-principles calculation.

To further determine the effective moment of this magnetic polarized state, we probe the state by applying a strong external magnetic field ($|B|\gg|B_C|$) along the c-axis with a nonmagnetic tip. When the magnetic state is polarized with an applied field, the state will always shift to the same energy direction regardless of the relative field orientation[13] (top inset schematic in Fig. 3**b**), which was also experimentally observed (Fig. 3**b**). Moreover, the positive energy shift indicates the state has a negative effective moment, calculated to be -5$\mu_B$ (or a Lande g factor of 10) based on the shift rate as 0.275meV/T (right inset of Fig. 3**b**). This large value is beyond the spin Zeeman effects (~1$\mu_B$) and indicates the additional polarized orbital magnetization. The anomalous Zeeman effect with an unusual moment or g factor has been observed in the electronic bands of kagome magnets[11,13], which is often linked to the Berry phase physics associated with magnetism and spin-orbit coupling[11,13,28,29].

Finally, we probe the local impurity-impurity interaction through extensive imaging and spectroscopy investigation. In Fig. 4**a**, we present the evolution of the impurity bound state with increasing perturbation strength caused by a second nearby impurity. We find that with decreasing spatial separation, the bound state progressively decreases in intensity and finally splits into two sub-peaks. Figure. 4**b** further compares three cases with one isolated impurity, two neighboring impurities, and three neighboring impurities, respectively. We find the quantized number of split impurity states matches with the coupled impurity number, highlighting their atomic-scale quantum-level coupling. Differential conductance maps at these corresponding splitting energies demonstrate their striking orbital hybridizations (Fig. 4**c**). For two neighboring impurities, the dI/dV maps clearly show a bonding ($\sigma$) and antibonding ($\sigma^*$) molecular orbital formation[4], which is consistent with the quantum coupling of two degenerate states. For three impurities, the dI/dV maps show the formation of one bonding ($\sigma$) and two antibonding ($\sigma_1^*$, $\sigma_2^*$) molecular orbitals, an unusual situation in which we will discuss below. From the apparent lattice geometry, the three neighboring impurities have $C_{3V}$ symmetry, which would have a doubly degenerate state $\sigma^*$ protected by the mirror symmetry[7]. The mirror symmetry operation, however, would transfer spin-up to spin-down. Therefore, the combination of ferromagnetic spin-polarization and spin-orbit coupling naturally breaks the mirror symmetry (i.e. $C_{3V}$ is broken into $C_3$), leading to the splitting of $\sigma^*$. Such splitting is analogous to splitting the bulk magnetic Weyl fermion line[12,19], which is protected by the crystalline mirror symmetry. In $Co_3Sn_2S_2$, the splitting energy of the Weyl fermion line[12] (or magnetic nodal line gap) is around 50meV, which is remarkably consistent with the energy splitting of $\sigma^*$ in Fig. 4**b**, attesting to their similar origin from the spin-orbit coupled magnetic ground state.

In conclusion, we report the first STM studies of the doped nonmagnetic impurity behavior in a correlated kagome magnet. Associated with the In impurities, we find an intense spin-orbit polarized bound state with an unusual magnetic moment and quantized energy splitting under impurity-impurity interaction. Our characterizations of the spin-orbit polarization and impurity-impurity interaction provide microscopic insight into the interplay between impurity geometry, magnetism, and spin-orbit coupling. Our atomic-scale discovery can contribute to understanding the macroscopic effects of In doping including the suppression of bulk magnetism and a metal-insulator-metal phase transition[20,30], which calls for further theoretical investigation of emergent nonmagnetic impurity behavior in a correlated kagome magnet. The spin-orbit polarized quantum impurity involves multiple degrees of freedom, including charge, spin, and orbital, the quantum control of which can open up a new pathway to design miniaturized data storage and spintronic devices for the implementation of quantum computing.

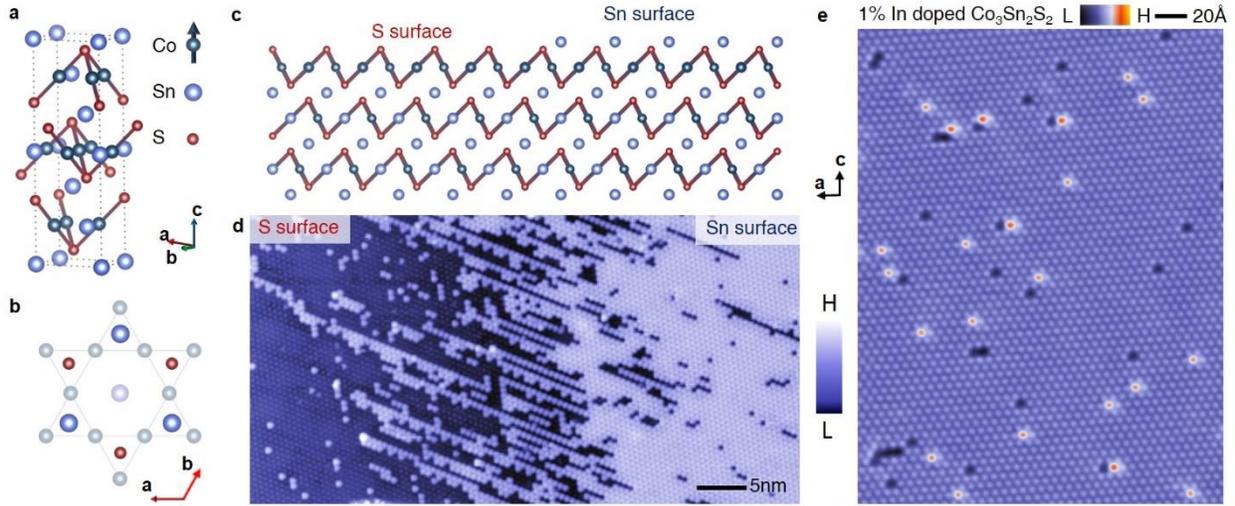

**Figure 1. Visualization of single atomic impurity in a spin-orbit coupled kagome magnet. a,** Crystal structure of $Co_3Sn_2S_2$. **b,** Hexagonal lattice of Sn layer and S layer with respect to the underlying Co kagome lattice. **c,** Side view of the crystal structure and illustration of the two possible terminating surfaces. **d,** Atomically resolved topographic image of the boundary between S surface and Sn surface. The Sn surface smoothly evolves into S surface with increasing coverage of Sn adatom. **e,** Atomically-resolved topographic image of Sn layer of 1% In doped $Co_3Sn_2S_2$.

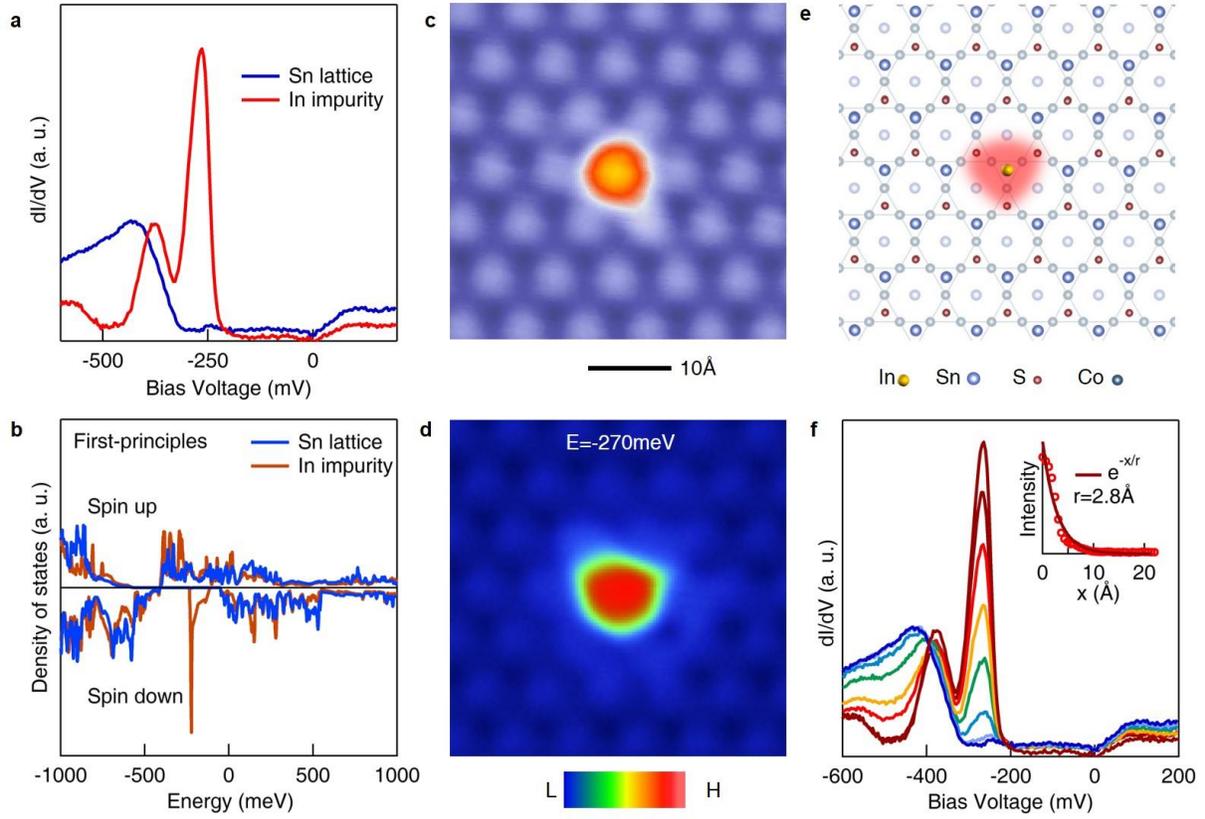

**Figure 2. Emergence of pronounced impurity bound state. a,** Differential conductance spectrums taken on the Sn lattice (blue) and at the In impurity (red), respectively. **b,** First-principles calculation of the spin-resolved local density of states of an In impurity and for the Sn lattice, which shows a magnetic impurity resonance. The calculation utilizes a 2×2×2 slab model with one In impurity, and the local density of states integrated from local In (or Sn), S and Co atoms. **c,** Topographic image of an isolated impurity. **d,** Corresponding differential conductance map taken at E = -270meV (resonance energy). **e,** Correlation between the atomic structure and the pattern in the differential conductance map. **f,** Differential conductance spectra taken across the surface with spatial variation from the center of the In impurity (dark red) to far away (blue).

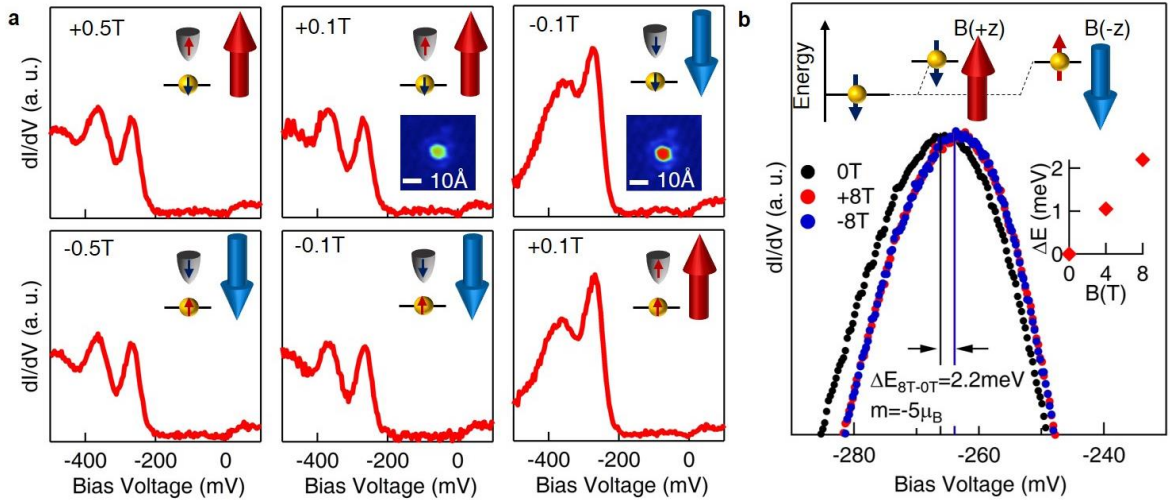

**Figure 3. Magnetic nature of the impurity bound state. a,** Dependence of the impurity state with Ni tip under a weak magnetic field. We apply +0.5T, +0.1T, -0.1T, -0.5T, -0.1T, +0.1T fields to systematically flip the magnetization of the tip and sample. The inset schematics illustrate the respective spin of the tip and the impurity state. Note the spin of the impurity state is anti-aligned with the bulk magnetization direction. Insets: respective differential conductance map taken at E = -270meV (resonance energy). **b,** Dependence of the impurity state with a strong magnetic field. Under both +8T and -8T, the peak exhibits a Zeeman energy shift of 2.2meV, which amounts to an effective moment of -5$\mu_B$. The inset shows the energy shift for different magnetic field magnitudes. Inset schematic illustrates the magnetization-polarized Zeeman effect. The applied field aligns the spins in the same orientation, hence +z and -z orientation fields leads the energy to shift in the same direction.

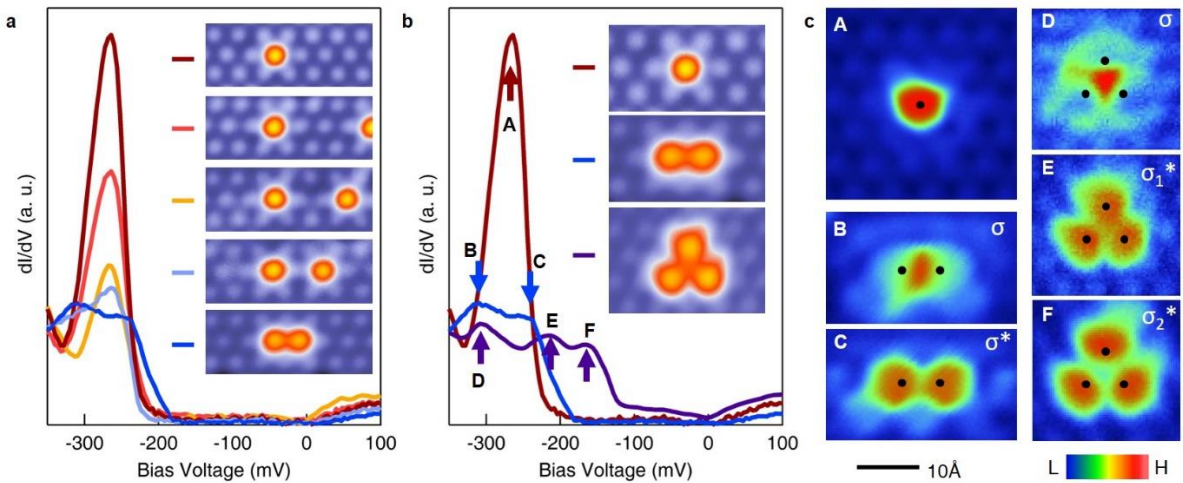

**Figure 4. Impurities induced quantized molecular orbitals. a,** Differential conductance spectra taken at the central impurity with perturbations of varying strengths from a second impurity. Inset: respective topographic images for impurity configurations. **b,** Local impurity state with coupling to different numbers of impurities. The arrows highlight the quantized splitting with additional

interacting impurity numbers. **c,** dI/dV maps at the respective bound state energies in **b**. dI/dV maps taken at A: -270meV for a single impurity; B, C: -315meV (bonding state σ), 240meV (antibonding state σ) for a double impurity, respectively; D, E, F: -310meV (bonding state σ), -220meV (antibonding state $\sigma_1^*$) and -170meV (antibonding state $\sigma_2^*$) for a triple impurity, respectively. The energy splitting of $\sigma_1^*$ and $\sigma_2^*$ is caused by ferromagnetic spin-polarization and atomic spin-orbit coupling. The black dots mark the center of impurities.

We thank Z. Song, T. Neupert and B. Lian for insightful discussions.